\documentclass[reprint,twocolumn,showpacs,preprintnumbers,superscriptaddress,amsmath,amssymb]{revtex4-2}
\usepackage{graphicx}
\usepackage{siunitx}
\usepackage{xcolor}
\usepackage{amsmath,amssymb}
\usepackage[T1]{fontenc}
\usepackage{lmodern}
\usepackage[utf8]{inputenc}
\usepackage[english]{babel}
\usepackage{natbib}
\usepackage{hyperref}
\hypersetup{colorlinks=true,citecolor={blue},linkcolor={blue},urlcolor={blue}}
\usepackage{siunitx}
\newcommand{\um}{\text{{\textmu}m}}

\begin{document}
\title{Hybridization of Non-Hermitian  Topological Interface Modes}

\author{Yuhao Wang}
\affiliation{School of Electrical and Electronic Engineering, Nanyang Technological University, Singapore 639798}
\affiliation{Institute of Materials Research and Engineering, Agency for Science Technology and Research (A*STAR), 2 Fusionopolis Way, Singapore 138634}

\author{Hai-Chau Nguyen}
\affiliation{Naturwissenschaftlich–Technische Fakult\"{a}t, Universit\"{a}t Siegen, 57068 Siegen, Germany}

\author{Zhiyi Yuan}
\affiliation{School of Electrical and Electronic Engineering, Nanyang Technological University, Singapore 639798}
\affiliation{Institute of Materials Research and Engineering, Agency for Science Technology and Research (A*STAR), 2 Fusionopolis Way, Singapore 138634}
\affiliation{ CNRS-International-NTU-Thales ResearchAlliance (CINTRA), IRL 3288, Singapore 637553}

\author{T. Thu Ha Do}
\affiliation{Institute of Materials Research and Engineering, Agency for Science Technology and Research (A*STAR), 2 Fusionopolis Way, Singapore 138634}

\author{Vytautas Valuckas}
\affiliation{Institute of Materials Research and Engineering, Agency for Science Technology and Research (A*STAR), 2 Fusionopolis Way, Singapore 138634}

\author{Hai Son Nguyen}
  \email{hai-son.nguyen@ec-lyon.fr}
\affiliation{Ecole Centrale de Lyon, INSA Lyon, Universit\'e Claude Bernard Lyon 1, CPE Lyon, CNRS, INL, UMR5270, 69130 Ecully, France}
\affiliation{Institut Universitaire de France (IUF), F-75231 Paris, France}

\author{Cuong Dang}
\email{hcdang@ntu.edu.sg}
\affiliation{School of Electrical and Electronic Engineering, Nanyang Technological University, Singapore 639798}
\affiliation{ CNRS-International-NTU-Thales ResearchAlliance (CINTRA), IRL 3288, Singapore 637553}

\author{Son Tung Ha}
\email{ ha\_son\_tung@imre.a-star.edu.sg}
\affiliation{Institute of Materials Research and Engineering, Agency for Science Technology and Research (A*STAR), 2 Fusionopolis Way, Singapore 138634}


\begin{abstract}

We propose and experimentally demonstrate the hybridization of radiating topological interface states, analogous to Jackiw-Rebbi states but in gain media with radiation fields. This hybridization not only modifies energy levels under a strong coupling scheme but also significantly reshapes far-field radiation characteristics. The bonding mode exhibits sub-radiant, omnidirectional emission, while the antibonding mode becomes super-radiant and highly unidirectional. Crucially, this non-Hermitian hybridization is tunable, allowing simultaneous control of energy splitting, quality factor, and far-field radiation by varying the distance between the two topological interfaces. Our findings establish hybridized radiating topological interface states as a robust platform for engineering two-level systems with tailored far-field responses, offering new possibilities for applications in beam shaping, nonlinear optics, quantum technologies, and beyond.

\end{abstract}

\maketitle

\emph{Introduction -} Topological photonics has recently emerged as a rapidly advancing and influential field, introducing novel paradigms for manipulating light through topologically protected  phenomena~\cite{ota_active_2020,khanikaev_photonic_2013,lu_topological_2014}. Drawing inspiration from the concept of topological insulators in condensed matter physics~\cite{fruchart_introduction_2013}, this field has paved the way for the design of optical structures with exceptional resilience and control. A defining feature of topological systems is the presence of robust edge states, also referred to as interface states, which are localized at the boundary between two photonic structures with different topological phases. These states are protected against defects and disorder, enabling a range of applications in waveguides, micro-resonators, plasmonic structures, and metasurfaces~\cite{lu_topological_2014}. 
At the heart of topological interface states lies the Jackiw-Rebbi (JR) state, an archetypal topological interface state originally introduced in a relativistic quantum model~\cite{jackiw_solitons_1976}. Derived from solutions to the Dirac equation, the JR model predicts robust zero-energy modes localized at scalar field transitions. This concept has since been extended to various domains, including condensed matter physics and photonics, where JR states emerge as topologically protected states at interfaces between systems with opposing topological invariants. 
The most common photonic realizations involve optical tight-binding lattices~\cite{blanco-redondo_topological_2016,Pan2018-rc,tran_extreme_2019,PhysRevA.96.013831,xu_plasmonic_2019,gorlach_photonic_2019}, where JR states localize at defects bridging dimer chains. 
More recently, JR states have been investigated in subwavelength lattices of multilayer thin films~\cite{gupta_realization_2024} and dielectric/plasmonic gratings~\cite{lee_topological_2021,choi_topological_2023,an_topological_2024}. These configurations enable the realization of JR states at interfaces of hetero-junction exhibiting band inversions. Notably, these systems leverage radiative leakage—a distinctive non-Hermitian feature of photonic platforms related to far-field radiation. In this context, JR states have been shown to exhibit polarization-independent behavior~\cite{gupta_realization_2024} and highly directional beaming~\cite{lee_topological_2022}, making them promising candidates for beam shaping, wavefront control, and other advanced photonic applications. However, these studies are restricted to single JR states, and the interplay between near-field confinement and far-field manipulation in more complex scenarios—such as the hybridization of JR states or higher-dimensional physics—remains largely unexplored.
\begin{figure}[htb!]
    \centering
    \includegraphics[width=0.7\linewidth]{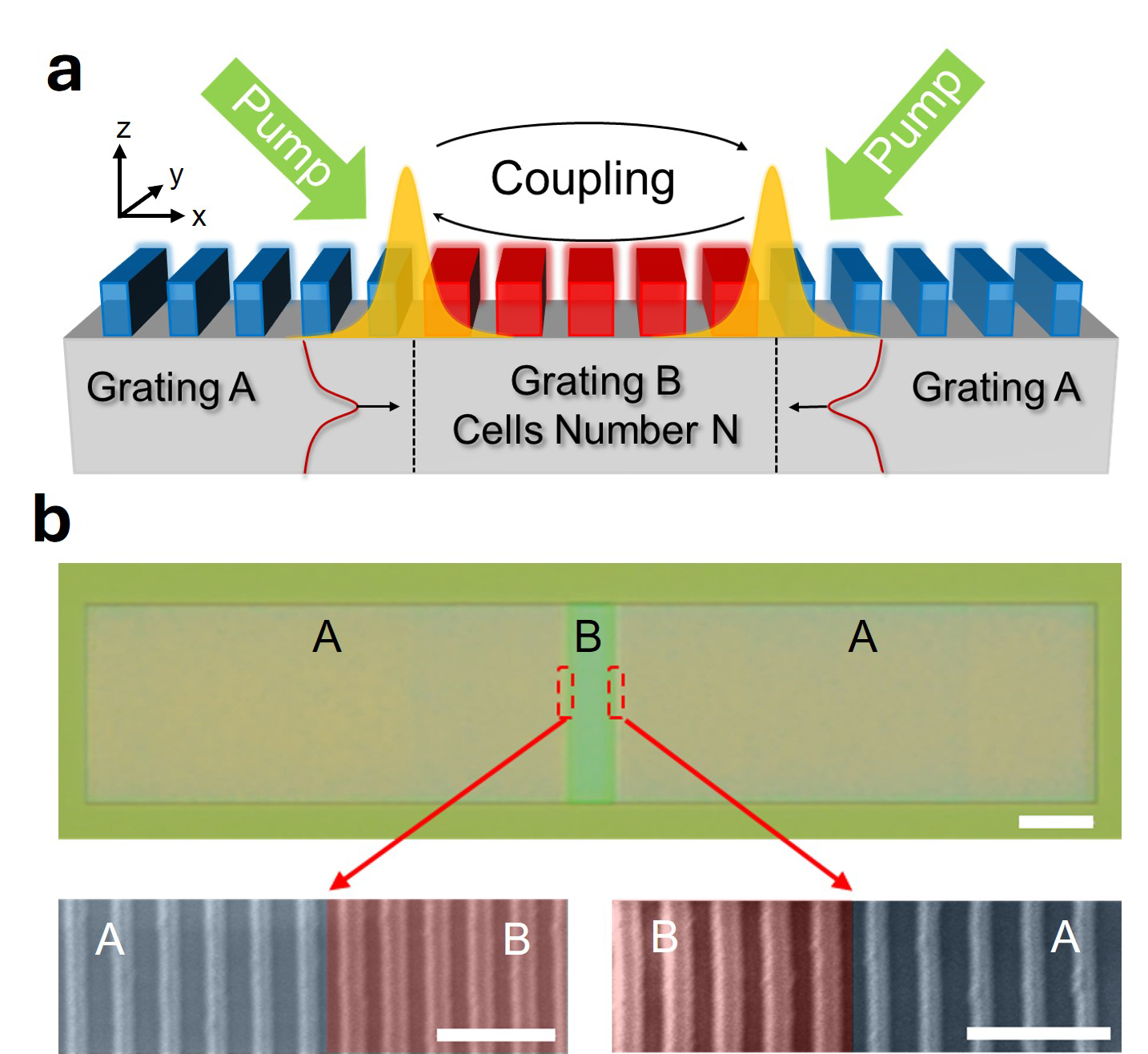}
    \caption{(a) Schematic representation of the double JR platform. Gratings A (blue), characterized by a negative Dirac mass, are positioned on either side of the array. Grating B (red), consisting of N periods, is placed at the center. This configuration supports two JR modes at the interfaces between gratings A and B. The spatial overlap between the two JR modes increases, leading to their coupling. (b)  Optical and SEM images of structure with N = 10. The scale bar are $5\um$ and $1\um$ in the optical and SEM image respectively.} 
    \label{Fig1}
    \end{figure}

In this Letter, we experimentally demonstrate the hybridization of two JR modes in an all-dielectric double-grating structure. By precisely tuning the separation between the two interfaces, their electric fields overlap and couple, leading to a pronounced energy splitting into bonding and antibonding modes. Notably, in such strong coupling regime, a striking radiative loss exchange occurs: the bonding mode becomes almost non-radiant, while the antibonding mode exhibits enhanced brightness and directionality. Integrating the structure with an emitting layer enables us to directly probe these effects through emission spectrum linewidth analysis. 
The non-radiant bonding mode, characterized by a narrow linewidth and high $Q$-factor, is promising for lasing applications, whereas the enhanced emission properties of the antibonding mode can be exploited for beam shaping and optical routing.

\emph{Design and Concept -} 
We first recall the band structure of sub-wavelength gratings and the mechanism responsible for generating the JR state in such systems. At the center of the Brillouin Zone (i.e., the $\Gamma$ point), the band structure of a photonic grating consists of pairs of bands, each pair comprising two bands with the same polarization (either transverse electric, TE, or transverse magnetic, TM) but opposite curvature in their band dispersion and distinct radiative leakage properties. One mode is highly leaky and is referred to as the “bright mode,” while the other mode is forbidden from radiating due to a symmetry mismatch between its field profile and the plane waves in the radiative continuum. This non-radiating mode is referred to as the “dark mode”~\cite{lee_band_2019,Lu2020}. 
The bandgap splitting between these two modes is determined by the filling fraction of the grating and can be freely tuned from negative to positive by adjusting the filling factor, enabling precise control over the band structure. This tuning corresponds to a band inversion, where the energy ordering of the two modes is reversed as the gap closes and then reopens~\cite{lee_band_2019,Ferrier2022}. This band inversion signifies a topological transition, and a JR state emerges at the double junction between two gratings whose band structures are inverted relative to one another~\cite{lee_topological_2021}.

In our design, to enable interaction between two JR states, two junctions are combined in an ABA configuration as shown in Figure~\ref{Fig1}(a). The gratings A and B are designed such that their band structures are inverted, resulting in JR states at the interfaces of the double-junctions AB and BA. Here, the finite grating B serves as the central separation grating, flanked by two semi-infinite gratings A. In this configuration, the interfaces AB and BA at the edges of grating B host two isolated JR states, which degenerate when the length of grating B is sufficiently large. Due to the finite size of array B (determined by the number of cells N), there is a non-zero overlap between the decaying tails of the two JR states, leading to the coupling between them. In the strong coupling regime, this coupling lifts the degeneracy, resulting in two hybridized JR states: the bonding mode and the antibonding mode. These terms are reminiscent of those used to describe diatomic molecules in atomic physics, where the symmetric and antisymmetric near-field profiles serve as distinguishing features of the two split modes. Moreover, the non-Hermitian nature of the JR states, due to their radiative leakage \cite{lee_topological_2022}, introduces a unique aspect to this hybridization that goes beyond the near-field interaction picture. Specifically, the far-field radiation of the JR states can interfere, corresponding to a loss exchange mechanism that modifies the radiative quality factors and far-field radiation patterns of the hybridized states compared to the uncoupled ones.

Theoretically, the photonic modes of this system are described by a two-component spinor $\Psi(x) = \begin{pmatrix} \Psi_+(x) \\ \Psi_-(x) \end{pmatrix},$
where the components $\Psi_\pm(x)$ represent the amplitudes of the forward and backward guided modes, respectively. The dynamics of this spinor are governed by a non-Hermitian Dirac Hamiltonian~\cite{sigursson_dirac_2024}:  
\begin{equation}\label{eq:H}
H = E_0 + \begin{pmatrix}
-iv \partial_x & m(x) \\
m(x) & iv \partial_x 
\end{pmatrix} -
i\left[\Gamma_{\rm nr} +\gamma \begin{pmatrix}
1 & -1 \\
-1 & 1
\end{pmatrix}\right],
\end{equation}
where $E_0$ is the midgap energy, and \( v \) represents the group velocity of the guided modes. The non-Hermitian terms include the nonradiative loss rate  \( \Gamma_{\rm nr} \) and the radiative loss rate \( \gamma \), which accounts for both the radiative decay of each guided mode into the continuum and the coupling between the two modes via the continuum. Importantly, the local Dirac mass term \( m(x) \) corresponds to the diffractive coupling strength and is responsible for the bandgap opening. By tuning the filling factor of the grating, the coupling strength \( m(x) \) can be continuously adjusted from negative to positive, inducing a band inversion~\cite{Lu2020,Ferrier2022}. For our system, the filling factors of gratings A and B are chosen such that the mass profile is defined as \( m(x) = m_B > 0 \) for \( 0 \leq x \leq d \) and \( m(x) = m_A < 0 \), with \( |m_A| \approx |m_B| \). This design ensures the presence of the JR states at the AB and BA interfaces. By solving for the quasi-bound states of Eq.\eqref{eq:H}, we can derive the energy, quality factor, as well as the near-field and far-field patterns of the hybridized JR states.
\begin{figure}[htb!]
    \centering
    \includegraphics[width=1\linewidth]{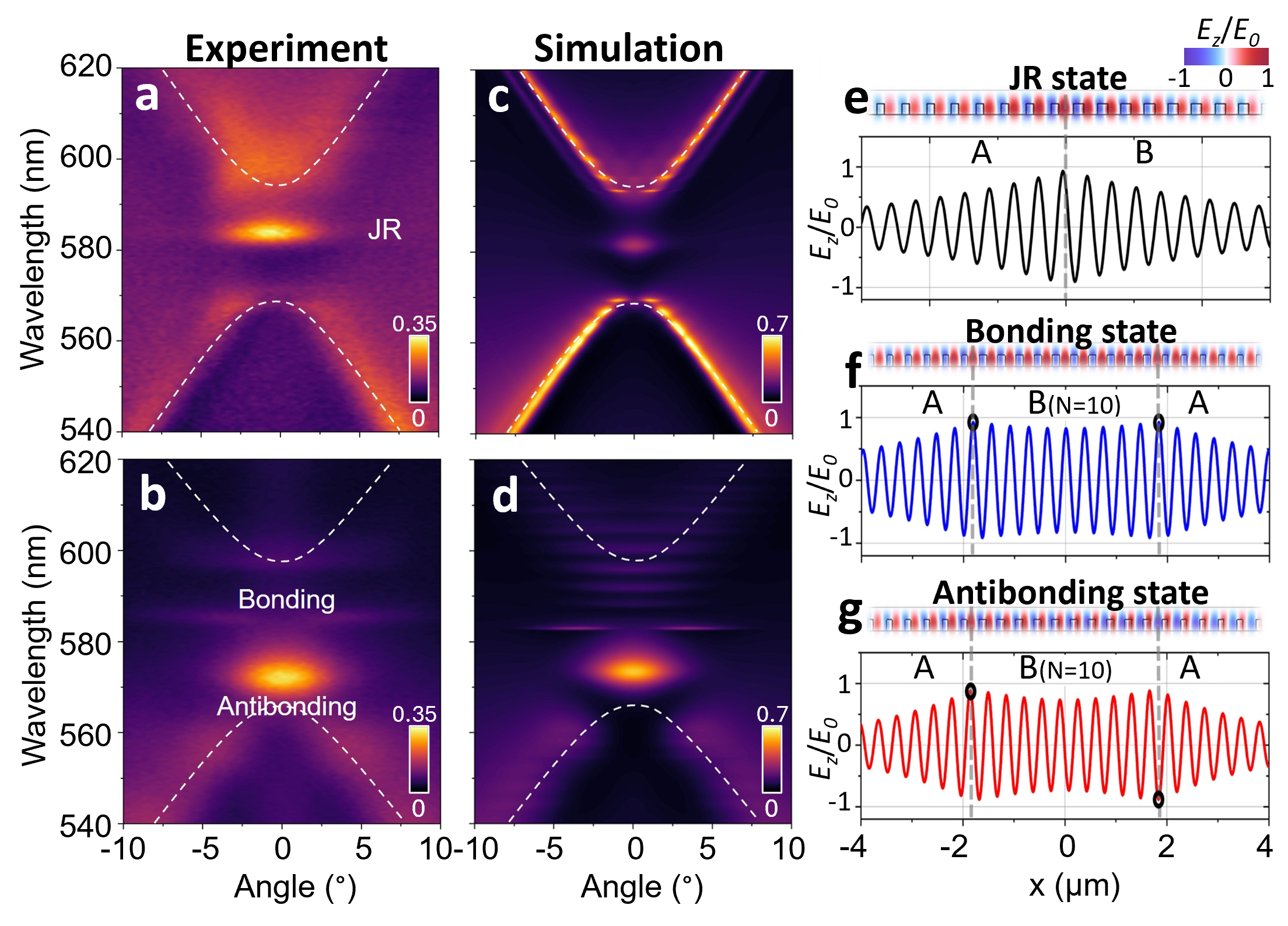}
    \caption{Experimental angle-resolved reflection spectra of the JR (a) and double-junction (b). (c-d) Simulated angle-resolved reflection spectra of the JR (c) and double-junction  (d). (e-f) Near-field distributions of the JR (e) and hybrid modes (f). Here, the electric field distributions presented in the pseudo-colormaps are calculated using FEM, while the oscillating functions are derived from our analytical model. The gray dashed lines indicate the interfaces of the normal JR junction and double-junction.}
    \label{Fig2}
\end{figure}

\emph{Observation of the JR state hybridization-} The grating structures are made of amorphous TiO\(_2\) (see  Figure~\ref{s1})  on a quartz substrate and fabricated via conventional E-beam lithography and dry-etching approach (see Appendix). Figure~\ref{Fig1}(b) presents the optical and SEM images of the double-junction structure, specifically for grating B with N = 10. Similar data for the single JR junction sample can be found in  Figure~\ref{s2}. The design of the grating structures is optimized so that the JR states are generated via the band inversion of the first two TM modes. To characterize the samples, we employ a home-built back focal plane (BFP) microspectrometer setup \cite{ha_directional_2018,Cueff2024}. Figure~\ref{Fig2}(a) shows the measured angle-resolved reflection spectra for the single JR junction sample with TM polarization. Here, the JR mode can be observed at ~580 nm, which is the midpoint of the bandgap of the two gratings A and B. This corresponds to a zero-energy interface state as predicted by our model (see Appendix) and is consistent with previous single JR reports for the grating systems \cite{lee_topological_2021,lee_topological_2022,an_topological_2024}. The band dispersion simulations and measured angle-resolved reflection spectra of the individual gratings A and B can also be found in Figure~\ref{s3}. 

The double-junction sample is measured by collecting the reflection signal from the whole central area (grating B) to characterize the hybridization of the double JR modes. As shown in Figure~\ref{Fig2}(b), the JR mode at the midgap region now disappears and is replaced by two other modes in the reflection spectra: a dark mode located at a longer wavelength (i.e., 583 nm) and a bright mode located at a shorter wavelength (i.e., 572 nm) compared to that of the JR mode. Figures~\ref{Fig2}(c,d) show the numerical simulation results obtained from Element Method (FEM) using commercial software (COMSOL MultiPhysics) for single JR and double-junction structures, respectively, which are in good agreement with the experimental observation in Figures~\ref{Fig2}(a,b). 
\begin{figure}[htb!]
    \centering
    \includegraphics[width=1\linewidth]{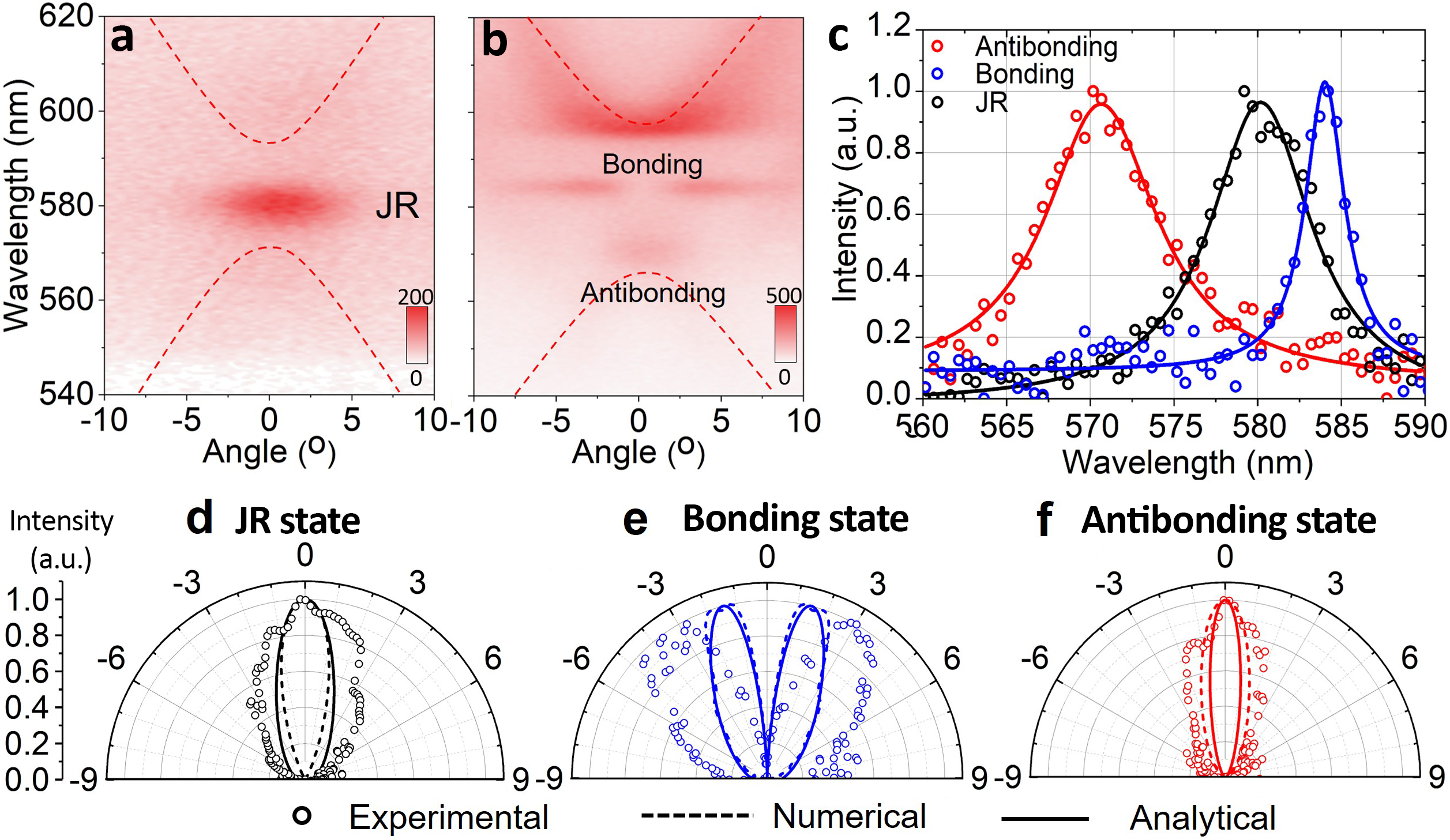}
    \caption{(a-b) Experimental angle-resolved PL spectra of the JR (a) and hybrid modes (b). (c) Comparison of PL spectra at a specific angle for different modes ($0^\circ$ for JR and antibonding modes; $3^\circ$ for bonding mode ). (d-f) Emission directionality of JR, bonding, and antibonding modes, respectively. Scattered dots, dash-dotted lines and solid lines represent experimental data, numerical simulations and analytical predictions respectively. }
    \label{Fig3}
\end{figure}
To further understand the formation of those modes, we study their near-field distributions using both numerical simulation and our analytical model. Figure~\ref{Fig2}(e) presents the near-field distribution of JR, with the gray dashed line marking the single-junction interface. The results show that the electric field component $E_z$  calculated from our analytical model is in good agreement with that extracted from the FEM simulation. For the case of double-junction structure, the near-field distribution exhibits distinct symmetry features for the two modes as shown in Figure~\ref{Fig2}(f). For the bright mode at the shorter wavelength, the electric field amplitude $E_z$ shows opposite signs at the two interfaces {(see marked points in the lower panel of Figure~\ref{Fig2}(f))}, with a zero value at the midpoint, indicating an asymmetric distribution. This mode is referred to as the antibonding mode. In contrast, the dark mode at the longer wavelength displays a symmetric electric field profile, where maximum $E_z$ is observed at both the interfaces and the midpoint {(see marked points in the upper panel of Figure~\ref{Fig2}(f))} corresponding to the bonding mode. The hybridization behavior of the two JR modes considered here resembles the atomic theory of bonding and antibonding hybridization and can be utilized as a platform to study light-matter interaction.

\emph{Emission of hybridized JR states--}
To study the emission behavior of the JR and their hybrid modes, dye molecules (Nile red, see photoluminescence (PL) spectrum in Figure~\ref{s4}) are mixed with PMMA and spin-coated onto the samples. To excite the dye, 532 nm laser is aligned to the above-mentioned BFP microspectrometer and focused onto the sample via the microscope objective. Figure~\ref{Fig3}(a) shows angle-resolved PL spectra of the dye molecules when coupled to the JR mode. In this case, the excitation laser is tightly focused onto the JR junction to minimize the background emission. It is noted that the PL is significantly enhanced compared to the background emission and well-coupled to the JR mode. This is because the mode profile extends from the TiO\(_2\) gratings into the PMMA/dye-filled region (Figure~\ref{Fig2}(e)), resulting in efficient spatial overlapping between the optical mode and emitter medium. For the case of the double-junction array, the laser spot is enlarged to cover both junctions. Thus, the PL can be seen coupling to both hybridization modes (i.e., bonding and antibonding) and the trivial modes of the array B as shown in Figure~\ref{Fig3}(b).

To further study the emission linewidth of the JR and hybrid modes, we extracted the PL spectra at the peak emission angle of each mode (i.e., $0^\circ$ for JR and antibonding modes and $3^\circ$ for bonding mode) as presented in Figure~\ref{Fig3}(c). By analyzing the PL spectra, the linewidth of the bonding and antibonding modes are determined to be 5 and 9 nm, respectively, compared to 8 nm of the JR mode. The observation in the linewidth variation in the hybrid modes reveals that the antibonding mode is more radiative (super-radiant), while the bonding mode is less radiative (sub-radiant) compared to the JR mode. This observation aligns well with our numerical simulation results shown in  Figure~\ref{s5}. Importantly, these findings confirm the exchange of radiative loss via mode hybridization, as predicted by our analytical model. It is worth noting that the total linewidths of our eigenmodes exhibit a significant contribution from non-radiative losses, which are not subject to the loss exchange mechanism. When considering only radiative losses, the linewidth of the antibonding mode can be an order of magnitude broader than that of the bonding mode (see Figure~\ref{fig:A1}).

The substantial alteration of radiative leakage due to loss exchange leads to a significant modification of the radiation patterns compared to that of a single JR. To explore this effect, we analyzed the emission directionality for each resonant mode. Figure~\ref{Fig3}(d) shows far-field emission pattern of JR mode, which is highly directional to the normal angle (i.e., within $3.5^\circ$), in good agreement with the previous results~\cite{lee_topological_2022,an_topological_2024}. Remarkably, upon hybridization of the JR modes, we observed that the antibonding mode exhibits even more directional emission (i.e., within  $2.5^\circ$) compared to that of the JR mode as shown in Figure~\ref{Fig3}(e). In contrast, the bonding mode shows a unique emission pattern where the radiation at the zero wavevector is significantly suppressed due to its dark mode nature, resulting in peak emission at an off-normal angle as shown in Figure~\ref{Fig3}(f).
\begin{figure}[htb!]
    \centering
    \includegraphics[width=1\linewidth]{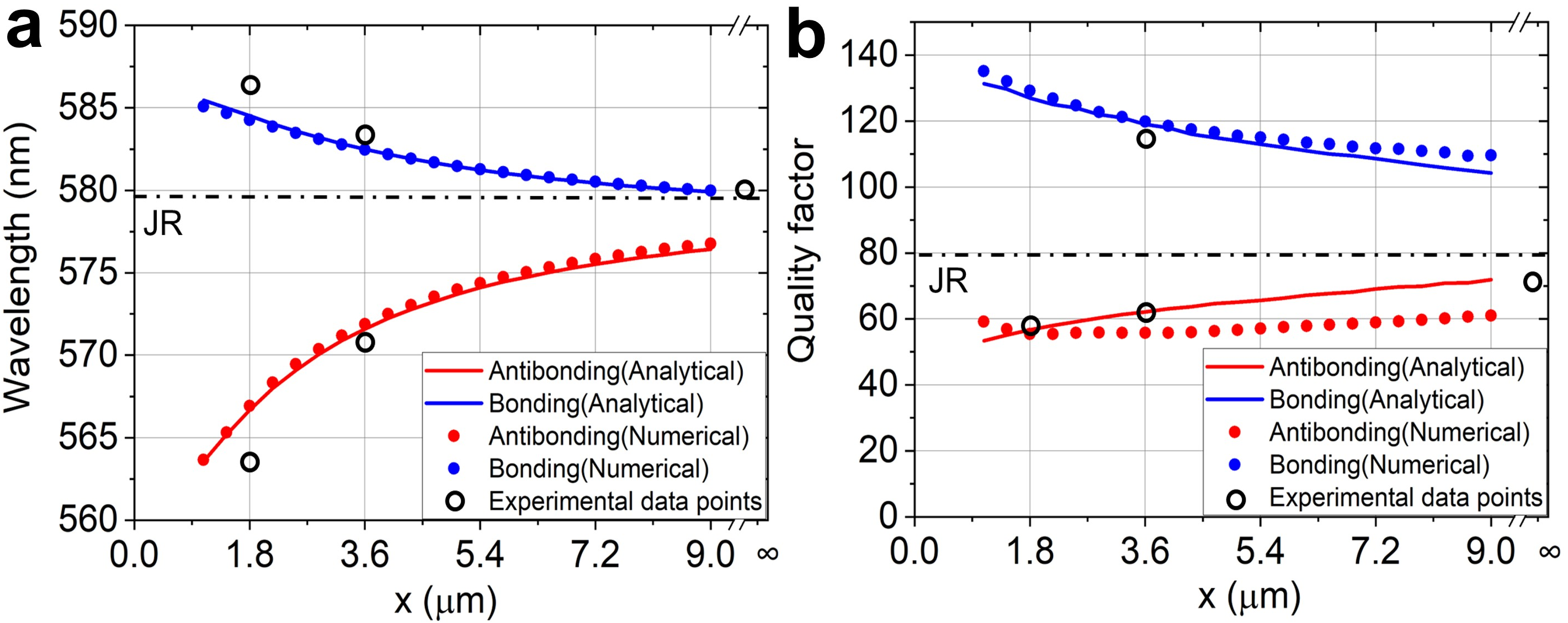}
    \caption{Evolution of the hybrid modes (a) and their Q-factor (b) when changing the size of grating B, showing good agreement between our analytical model, numerical simulation and experiment results.} 
    \label{Fig4}
\end{figure}

\emph{Tunability of the JR hybridization--} Adjusting the size of the middle array (i.e., array B) allows for on-demand control of the coupling strength.  When the separation distance is sufficiently large, the JR modes remain uncoupled, and no energy splitting occurs. As the separation distance decreases, the increased spatial overlap between JR modes (see Figure~\ref{s6}) leads to mode hybridization, resulting in the formation of bonding and antibonding modes. Further reducing the separation distance strengthens the coupling, leading to greater mode splitting (see Figure~\ref{s7}). When array B becomes sufficiently small, it no longer supports guided-mode resonances, and the interface mode disappears.  The splitting behavior observed in our analytical calculations closely matches numerical simulations and experimental results, as shown in Figure~\ref{Fig4}(a). Notably, for small array B sizes (e.g., $N = 5$), the loss exchange mechanism becomes highly efficient, rendering the bonding mode nearly non-radiant and barely detectable in our experiment (see Figure~\ref{s8}). The strong agreement between experimental results, numerical simulations, and analytical calculations further validates the accuracy of our theoretical model.  

To emphasize control over the loss exchange mechanism, we extracted the Q-factors of the hybrid modes as a function of the separation distance (Figure~\ref{Fig4}(b)). The Q-factor of the bonding mode increases rapidly with decreasing separation distance, indicating stronger mode coupling. In contrast, the antibonding mode exhibits a much weaker Q-factor variation due to its inherently leaky nature. By eliminating optical loss in the emitting layer, our simulations predict that the radiative Q-factor of the bonding mode can be significantly enhanced, reaching values up to $\sim10^4$ under the strongest coupling conditions (see Figure~\ref{fig:A1} in the Appendix). This result is particularly promising for applications requiring high optical confinement, such as lasing and optical sensing. Additionally, implementing dynamic tuning mechanisms—such as phase-change materials or electrically induced refractive index modifications in array B—could enable applications in optical switching and beam steering.

\emph{Conclusion--} Our work provides the first experimental demonstration of hybridization between topological interface  modes in non-Hermitian photonic systems, marking a significant advancement in the field of topological photonics. By employing an all-dielectric double-grating design, we have directly observed the mode splitting into bonding and antibonding states, which not only alters the energy landscape but also triggers a strong exchange of radiative losses between the coupled interfaces. This loss exchange manifests in pronounced modifications of both the radiative linewidths and the far-field radiation patterns of the hybrid modes. 
The ability to control the coupling strength—by adjusting the separation between the gratings—enables precise tuning of the energy splitting and the associated $Q$-factors. In this regime, the bonding mode, characterized by its dark mode nature and high $Q$-factor, offers enhanced light trapping, which is advantageous for lasing and sensing applications. Conversely, the antibonding mode exhibits super-radiant emission with highly directional far-field properties, paving the way for advanced beam shaping, optical routing, and dynamic beam steering applications. Furthermore, the observed long-range coupling distance ($\sim$3~\um) not only surpasses that found in conventional plasmonic systems \cite{domina_bonding_2020,huang_abnormal_2020,pal_angle-resolved_2023} but also rivals the best performance reported in nanophotonic and molecular platforms \cite{hoang2024collectivenaturehighqresonances,li_structural_2021}. Such extended light-matter interaction is critical for the development of large-scale integrated photonic circuits, where precise mode control and long-distance coupling are essential.
Beyond the immediate practical applications, our study establishes a robust theoretical framework for understanding and predicting the behavior of hybridized JR states in complex photonic environments. This framework, which leverages non-Hermitian physics to capture the essential dynamics of radiative loss exchange, provides a foundation for exploring more intricate systems, such as multi-mode hybridization and higher-dimensional topological phenomena.\\

\emph{Acknowledgements--}
The authors thank Ki Young Lee for fruitful discussions. S.T.H, T.T.H.D and V.V acknowledge funding support from A*STAR MTC-Programmatic Fund (M21J9b0085).
H.C.N was supported by the Deutsche Forschungsgemeinschaft (DFG, German Research Foundation, project numbers 447948357 and 440958198), 
the Sino-German Center for Research Promotion (Project M-0294), the ERC (Consolidator Grant 683107/TempoQ), the German Ministry of Education and Research (Project QuKuK, BMBF Grant No. 16KIS1618K),
and the EIN~Quantum~NRW. H.S.N acknowledges the financial support from the French National Research Agency (ANR) under the project POLAROID (ANR-24-CE24-7616-01).\\

\emph{Note added --} During the preparation of this manuscript, we became aware of the recent work by Choi \textit{et al.}~\cite{choi2024topologicalbeaminglightproofofconcept}, which experimentally investigates the far-field properties of a single JR state in a photonic grating. However, their study does not address the hybridization of JR states or the loss-exchange mechanism, which are central to and represent the novelty of our result.




\bibliography{ref}

\appendix

\renewcommand{\theequation}{A\arabic{equation}}
\renewcommand{\thefigure}{A\arabic{figure}}
\renewcommand{\thesection}{A\arabic{section}}

\setcounter{equation}{0}
\setcounter{figure}{0}
\setcounter{table}{0}

\section{Analytical model}
The eigenmodes in our grating structure are described by the dynamics of counter-propagating guided modes of a planar waveguide with an average refractive index.  The local periodic corrugation of the grating leads to a diffractive coupling of strength \( m \) between these guided modes and also couples them to the radiative continuum with a leakage rate \( \gamma \) . Moreover, the guided modes are also coupled via the radiative continuum (i.e., radiative coupling). In the case where the periodic corrugation is symmetric (which is the case in our work), this radiative-induced coupling strength is \( -\gamma \)~\cite{sigursson_dirac_2024}, where the `$-$' sign is due to the interference between radiations of opposite polarizations. 
Under the envelope function approximation~\cite{Bastard1981}, a spatial modulation of the corrugation gives rise to a modulation of the diffractive coupling strength \( m(x) \). This modulation is the key ingredient for generating JR states, which are localized at the interface where \( m(x) \) switches signs. As a consequence, the two-component spinor  $\Psi(x,t) = \begin{pmatrix} \Psi_+(x,t) \\ \Psi_-(x,t) \end{pmatrix}$, where the components \( \Psi_+(x,t) \) and \( \Psi_-(x,t) \) represent the amplitudes of the forward and backward guided modes, respectively, is governed by the dynamic equation (NB: here we adopt the convention $e^{-i\omega t}$ to describe temporal oscillation of electromagnetic wave) $\partial_t\Psi = -iH\Psi$. Here $H$ is a non-Hermitian Dirac Hamiltonian, given by:
\begin{widetext}
\begin{equation} \label{eq:Hamiltonian}
H(x) =   \underbrace{\begin{pmatrix}
U(x)- i v \partial_x & m(x) \\
m(x) & U(x) + i v \partial_x
\end{pmatrix}}_{\text{Hermitian Dirac Hamiltonian } H_D}
\quad
\underbrace{-i \gamma 
\begin{pmatrix}
1 & -1 \\
-1 & 1
\end{pmatrix}}_{\text{Radiative losses $-i H_R$}} 
\quad
\underbrace{-i\Gamma_{{\rm nr}}}_{\text{Non-radiative losses}}
\end{equation}
\end{widetext}
The offset energy \( U(x) \) corresponds to the mid-gap energy of the local band structure. Its spatial variation is small and can be neglected compared to that of \( m(x) \). However, we incorporate this variation in the model to quantitatively reproduce the numerical and experimental results. It can be considered as a local potential term.

Locally, the forward and backward waves corresponds are of wavectors $\pm 2\pi/a$ where $a$ is the local period of the grating that might be slowly modulated. 
With the counter-propagating core functions of $e^{\pm i \frac{2\pi}{a} x}$, the nearfield pattern is then given by~\cite{sigursson_dirac_2024} $\mathbf{E}^\text{nearfield}(x,z)=\Psi_+(x)\mathbf{E}_{+}(z) e^{i \frac{2\pi}{a} x} + \Psi_-(x)\mathbf{E}_{-}(z) e^{-i \frac{2\pi}{a} x} $ for the electric field of the two guided modes; and 
    $ \mathbf{H}^\text{nearfield}(x,z)=\Psi_+(x)\mathbf{H}_{+}(z) e^{i \frac{2\pi}{a} x} + \Psi_-(x)\mathbf{H}_{-}(z) e^{-i \frac{2\pi}{a} x}$ for the magnetic field. 

Since our guided modes are of TM (Transverse Magnetic) nature, their magnetic field has only $y$ component, while their electric field has $x$ and $z$ components \cite{Hunsperger2009}. Moreover a $\pi$ rotation around $z$ axis transforms $\mathbf{E}_+(z)$ and $\mathbf{H}_+(z)$ to $\mathbf{E}_-(z)$ and $\mathbf{H}_-(z)$ respectively, and vice versa. Therefore $\mathbf{E}_\pm(z)=\pm E_x(z)\mathbf{\hat{x}} + E_z(z)\mathbf{\hat{z}}$ and $
     \mathbf{H}_\pm(z)=\pm H_y(z)\mathbf{\hat{y}}$. As a consequence, at a given $z$ position, the nearfield patterns for different field component are give by:
\begin{equation}\label{eq:nearfield}
\begin{split}
       E^\text{nearfield}_x(x) &\propto\Psi_+(x) e^{i \frac{2\pi}{a} x} - \Psi_-(x) e^{-i  \frac{2\pi}{a} x}, \\
       E^\text{nearfield}_z(x) &\propto\Psi_+(x) e^{i \frac{2\pi}{a} x} + \Psi_-(x) e^{-i  \frac{2\pi}{a} x},\\
       H^\text{nearfield}_y(x) &\propto\Psi_+(x) e^{i \frac{2\pi}{a} x} - \Psi_-(x) e^{-i  \frac{2\pi}{a} x}.
\end{split}
\end{equation}

From the dynamic equation $\partial_t\Psi = -iH\Psi$ and the radiative losses of the non-Hermitian Hamiltonian given in Eq.~\eqref{eq:Hamiltonian}, we obtained the radiation flux (i.e. farfield radiation) as:
\begin{equation} \label{eq:farfield_x}
    I^\text{farfield}(x)=\gamma \Psi^\dagger H_R \Psi = \gamma \lvert{\Psi_{+}(x) - \Psi_{-}(x)}\rvert^2
\end{equation} 
The farfield pattern in momentum space is then given by:
\begin{equation} \label{eq;farfield_k}
    I^\text{farfield}(k)\propto \gamma \lvert{\tilde{\Psi}_{+}(k) - \tilde{\Psi}_{-}(k)}\rvert^2
\end{equation}
where $\tilde{\Psi}_{\pm}(k) \propto \int_{-\infty}^{+\infty}\Psi_{\pm}(x)e^{ikx}dx$ is the Fourier transform of $\Psi(x)$.\\ 

To obtain a double junction of JR states, the mass $m(x)$ in our system, consisted of a finite grating B sandwiched between two semi-inifite grating A, is given as $ m(x) = m_A < 0$ for $|x| > d/2$ and $m(x)=
m_B > 0$ for $|x| < d/2$. There is also a small variation of the offset energy, given by $U(x) = E_A$ for $|x| > d/2$ and $U(x)=E_B$ for $|x| < d/2$. The designs of the two gratings A and B have been optimized so that $|m_A|\approx|m_B|$ and $E_A \approx E_B$, therefore the two gratings exhibit well aligned bandgaps but inverted.  \\ 

In the approximation where the piecewise-varying mass profile $|m(x)|\gg \gamma,\gamma_{nr}$, one may first solve the Hermitian Dirac equation:
\begin{equation}
    H_D\Psi = E\Psi
\end{equation}
where $H_D$ is the Hermitian part of the Hamiltonian in Eq.\eqref{eq:Hamiltonian}. Applying transfer matrix method for Dirac equation~\cite{Nguyen_2009}, the energy $E$ the hybrid modes is given by the following equation:
\begin{equation}\label{eq:Ebis}
\coth \left( d \sqrt{\frac{m_B^2 - \Delta E_B^2}{v}} \right) = \frac{\Delta E_A\Delta E_B - m_A m_B}{\sqrt{m_A^2 - \Delta E_A^2} \sqrt{m_B^2 - \Delta E_B^2}}
\end{equation}
where $\Delta E_A=E-E_A$, $\Delta E_B=E-E_B$, with $E_A$ and $E_B$ being the mid-gap energy of the grating A and B respectively ($E_A\approx E_B$). 
By solving Eq.\eqref{eq:Ebis}, we obtain two solutions for energy \( E \), which correspond to the energy of the two hybrid JR states. For each eigenvalue, we also obtain the corresponding eigenvector \( \Psi(x) = (\Psi_+(x), \Psi_-(x)) \). The nearfield patterns of these states are then calculated using the Eq.\eqref{eq:nearfield}.\\

\emph{Chiral symmetry --} In the absence of the local potential, $E_A=E_B=E_0$,  the equation \eqref{eq:Ebis} gives two solutions with opposite signs with respect to the offset energy $E_0$ (i.e. if $\Delta_E$ is a solution, thus so is $-\Delta_E$). This reflects the chiral symmetry of the Hermitian Dirac Hamiltonian $H_D$ under the transformation by the Pauli matrix $\sigma_y$: $\sigma_y H_D \sigma_y = -H_D $. For a single JR junction, this chiral symmetry enforces the single topological edge state to be exactly at zero energy with respect to $E_0$. The energies of hybridized JR states are no longer at zero energy, but coming in pair of opposite signs. By introducing the non-zero local potential $U(x)$,  this  chiral symmetry is broken.

\emph{Parity symmetry --} In general, the ABA double heterojunction leads to an engineered mass profile $m(x)$ and a local potential $U(x)$ that satisfy $ m(x)=m(-x)$ and $U(x)=U(-x)$. Therefore, even in presence of a local potential $U(x)$ the Hamiltonian $H_D$ always exhibits a parity symmetry described by $\sigma_x P$, where $P$ transforms $x$ into $ - x$ and $\sigma_x$ is Pauli matrix that exchanges the spinor components $\Psi_+$ and $\Psi_{-}$ (i.e. swapping left-going wave with right-going wave): $ (\sigma_x P) H_D (\sigma_xP) = H_D$. This brings the inspiration to classify hybridized topological states into bonding  (i.e. even parity) and antibonding states  (i.e. odd parity), which are even or odd under spatial reflection, respectively, in similarity with the tight-binding theory of quantum chemistry. Therefore, for the Bonding mode, $E^\text{nearfield}_z(x)$ is an even function, while $E^\text{nearfield}_x(x), H^\text{nearfield}_y(x)$ are odd functions. On the other hand, for the Antibonding mode, $E^\text{nearfield}_z(x)$ is an odd function, while $E^\text{nearfield}_x(x), H^\text{nearfield}_y(x)$ are even functions. Moreover, using the Eq \eqref{eq:nearfield}, we deduce that: $\Psi_+(x)=\Psi_-(-x)$ for the Bonding mode, and  $\Psi_+(x)=-\Psi_-(-x)$ for the Antibonding mode. Interestingly, the previous conditions, applied to the farfield expression of Eq \eqref{eq;farfield_k} leads to $I^\text{farfield}(k=0) = 0$ for the Bonding mode. Such result explains the lack of radiation of the bonding mode at $k=0$, and its high radiative quality factor observed in the experiment and in the numerical simulations.\\

Once the bound states solution obtained, the radiative losses of hybrid JR states are calculated by applying the losses term $i\Gamma$ to the eigenvectors $\Psi$. For instance, the farfield radiation pattern in momentum space is calculated using the Eq.~\eqref{eq;farfield_k}, and the imaginary energy component corresponding to radiative losses is given by:
\begin{equation}
    \Gamma_{\rm r}= \frac{\int{\Psi^\dagger H_R \Psi}}{\int{\Psi^\dagger\Psi}} = \gamma\frac{\int_{-\infty}^{+\infty} \left| \Psi_+(x) - \Psi_-(x) \right|^2 \, dx}{\int_{-\infty}^{+\infty} \left( \left| \Psi_+(x) \right|^2 + \left| \Psi_-(x) \right|^2 \right) \,dx}
\end{equation}

Including both radiative and non-radiative losses, the total imaginary energy component is given by $\Gamma_{\rm tot}= \Gamma_{\rm nr} + \Gamma_{\rm r}$. Finally, the total quality factor $Q_{\rm tot}$ and the radiative quality factor $Q_{\rm r}$ can be calculated, using $Q=\frac{E}{2\Gamma_{\rm tot}}$ and $Q_{\rm r} =\frac{E}{2\Gamma_{\rm r}}$.\\

\begin{figure}[htb!]
    \centering
    \includegraphics[width=0.8\linewidth]{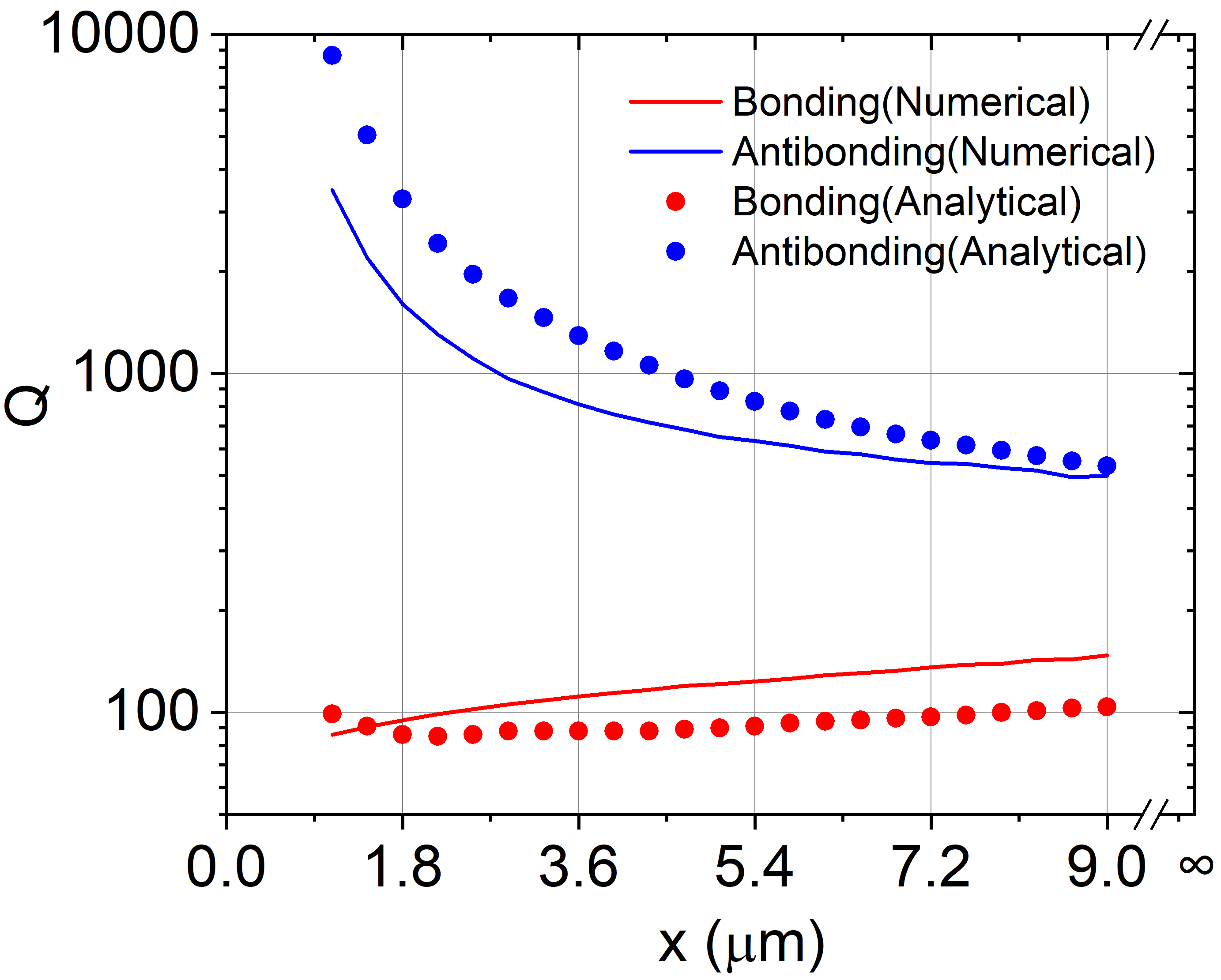}
    \caption{Evolution of Q factor respecting to the change of separation distance (grating B cells number N) without loss in emitting layer.}
    \label{fig:A1}
\end{figure}

\emph{Parameters--} Numerically, the parameters of our model can be extracted from numerical simulation of single grating A and B : the eigenvalues obtained from FEM simulations are fitted using the solution when $m(x)=m_{A,B}$ and $U(x)=E_{A,B}$. By doing that, for the design of single JR junction, we obtain the mass of grating A  \( m_A = -50 \, \text{meV} \); the mass of grating B \( m_B = 40 \, \text{meV} \); since the bandgap is perfectly aligned, the two off-set energies are the same  \( E_A = E_B = E_0 = 2134 \, \text{meV} \); the group velocity is  \( v = 114.5 \, \text{meV} \cdot \um \). For the double-junction case, there is shrinkage of the bandgap of grating B and the bandgaps are less aligned. Here, \( m_A = -50 \, \text{meV} \), \( m_B = 20 \, \text{meV} \), \( E_A = 2165 \, \text{meV} \), \( E_B = 2135 \, \text{meV} \), \( v = 114.5 \, \text{meV} \cdot \um \). Regarding the losses, we take non-radiative loss \(\Gamma_{\rm nr }=7.12\text{meV}\) and radiative loss \(\gamma_{\rm }=6.53\text{meV}\). The comparision between quality factors obtained by numerical simulations and the analytical models are shown in Fig.~\ref{Fig4}(b) and Fig.~\ref{fig:A1} for the case with and withoud nonradiative losses respectively.

\section{Fabrication}

\emph{Deposition--} A ~180 nm TiO\(_2\) layer is deposited on a quartz substrate using ion-assisted deposition (IAD, Oxford Optofab3000). The refractive index data of our TiO\(_2\) is shown in Figure~\ref{s1}. Subsequently, a ~30 nm Cr layer is deposited on top by an electron beam evaporator (Angstrom EvoVac) to reduce the charging effect during the E-beam lithography process and to act as a hard-mask layer in the subsequent etching step. 

\emph{E-Beam Lithography--} Hydrogen silsesquioxane (HSQ, EM Resist, 6\%) is used as a negative resist. The HSQ resist layer is spin-coated onto the sample at 3000 rpm for 60 seconds and subsequently baked at 180\(^\circ\)C for 3 minutes on a hotplate. Nanopatterning is done with E-beam lithography (Elionix ELS 7000, 100 keV, 500 pA). Afterwards, the HSQ patterns are developed by immersing the sample in a homemade salty developer (aqueous solution of 1 wt\% NaOH and 4 wt\% NaCl) for 4 minutes.

\emph{Etching--} HSQ patterns are transferred onto the Cr layer by inductively coupled plasma reactive ion etching (ICP-RIE, Oxford Plasmalab 100) with Cl\(_2\) gas. Subsequently, the patterns are transferred onto TiO\(_2\) layer using the Cr hard-mask by the same etching tool with CHF\(_3\) gas. Finally, the Cr hard-mask is removed by wet etching using chromium etchant solution (Merck) for 10 minutes.

\emph{Emitting Layer Preparation--} Dye solution (Nile Red, Sigma-Aldrich) mixed with 950K A4 PMMA to a concentration of 1 mmol/L is used as the emitting layer. The mixture is then spin-coated on the sample at 5000 rpm to form a ~200 nm thick film. A refractive index of 1.51 for PMMA is utilized in our theoretical calculations and numerical simulations.

\section{Optical Characterization}
All optical measurements are conducted through a home-built back focal plane (BFP) microspectrometer \cite{ha_directional_2018,Cueff2024}, consisting of an inverted microscope (Nikon Ti-U) coupled with a spectrograph (Andor SR-303i) via a custom-built lens system. The BFP of the collecting microscope objective (Nikon, 50$\times$, 0.55 numerical aperture (NA)) is projected onto the entrance slit of the spectrograph, opening at 100 \um. The optical signals are then projected onto a 2-dimensional electron-multiplying charged-coupled detector (EMCCD, Andor Newton 971) after passing a grating with a groove density of 150 lines/mm. The result is 2-dimensional data resolved in both wavelength and collecting angle, which corresponds to the NA of the microscope objective. In our setup, a pinhole is placed at the image plane in the collection path to limit the optical signal from the region of interest (e.g., JR interface). This configuration will ensure a high signal-to-noise ratio and avoid the background signal.  For reflection measurements, the sample is excited by a halogen lamp through a linear polarizer. For PL measurements, a supercontinuum laser (NKT, SuperK) with a tuning wavelength of 532 nm is used to excite the sample.

\setcounter{equation}{0}
\setcounter{figure}{0}
\setcounter{table}{0}

\renewcommand{\theequation}{S\arabic{equation}}
\renewcommand{\thefigure}{S\arabic{figure}}
\renewcommand{\bibnumfmt}[1]{[S#1]}
\renewcommand{\vec}[1]{\boldsymbol{#1}}

\onecolumngrid 
 \begin{center}
\Large{--- SUPPLEMENTAL INFORMATION ---}
\end{center}
\normalsize{}
\begin{figure*}[htb!]
    \centering
    \includegraphics[width=0.6\linewidth]{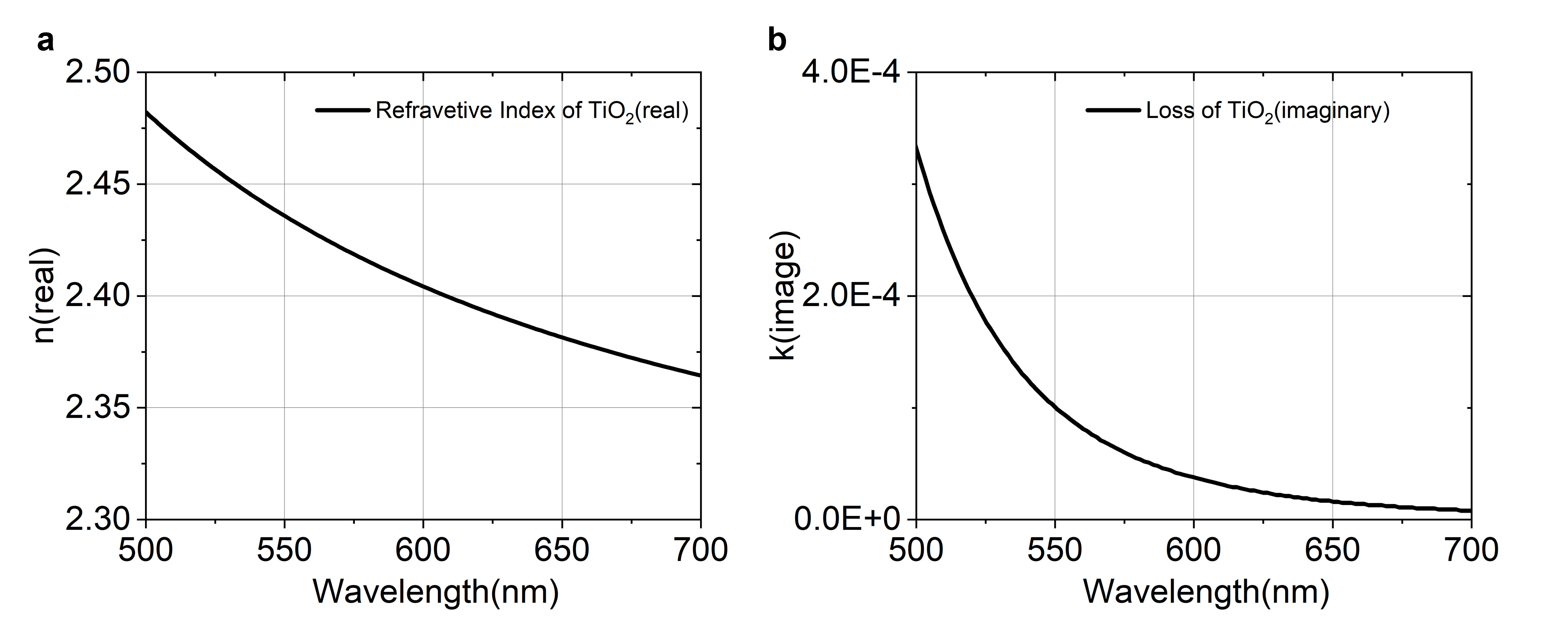}
    \caption{ Refractive index of TiO$_2$. (a) real part and (b) imaginary part.}
    \label{s1}
\end{figure*}

 \begin{figure*}[htb!]
    \centering
    \includegraphics[width=0.3\linewidth]{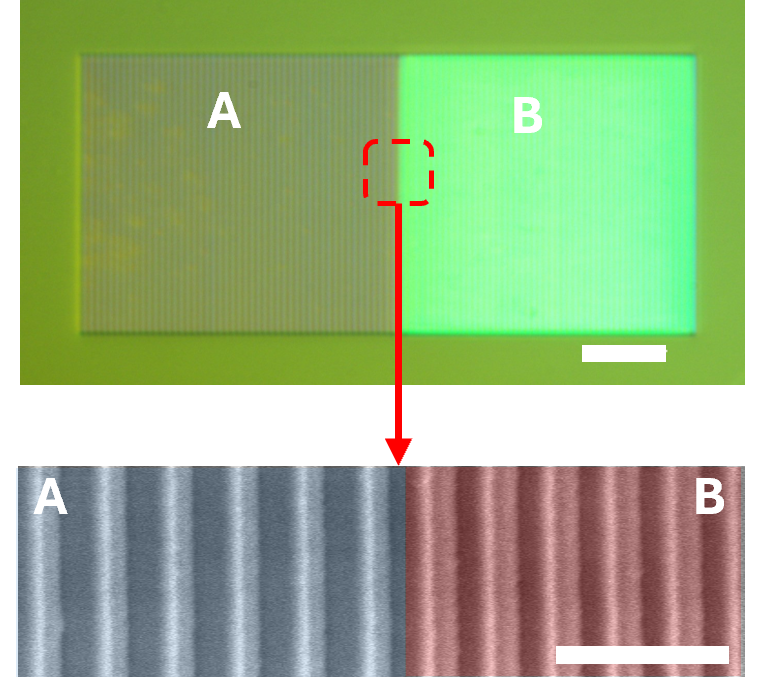}
    \caption{ Optical and SEM images of the single JR sample. The scale bar in the optical image is $5\um$; The scale bar in the SEM image is $1\um$.}
    \label{s2}
\end{figure*}

\begin{figure*}[htb!]
    \centering
    \includegraphics[width=0.7\linewidth]{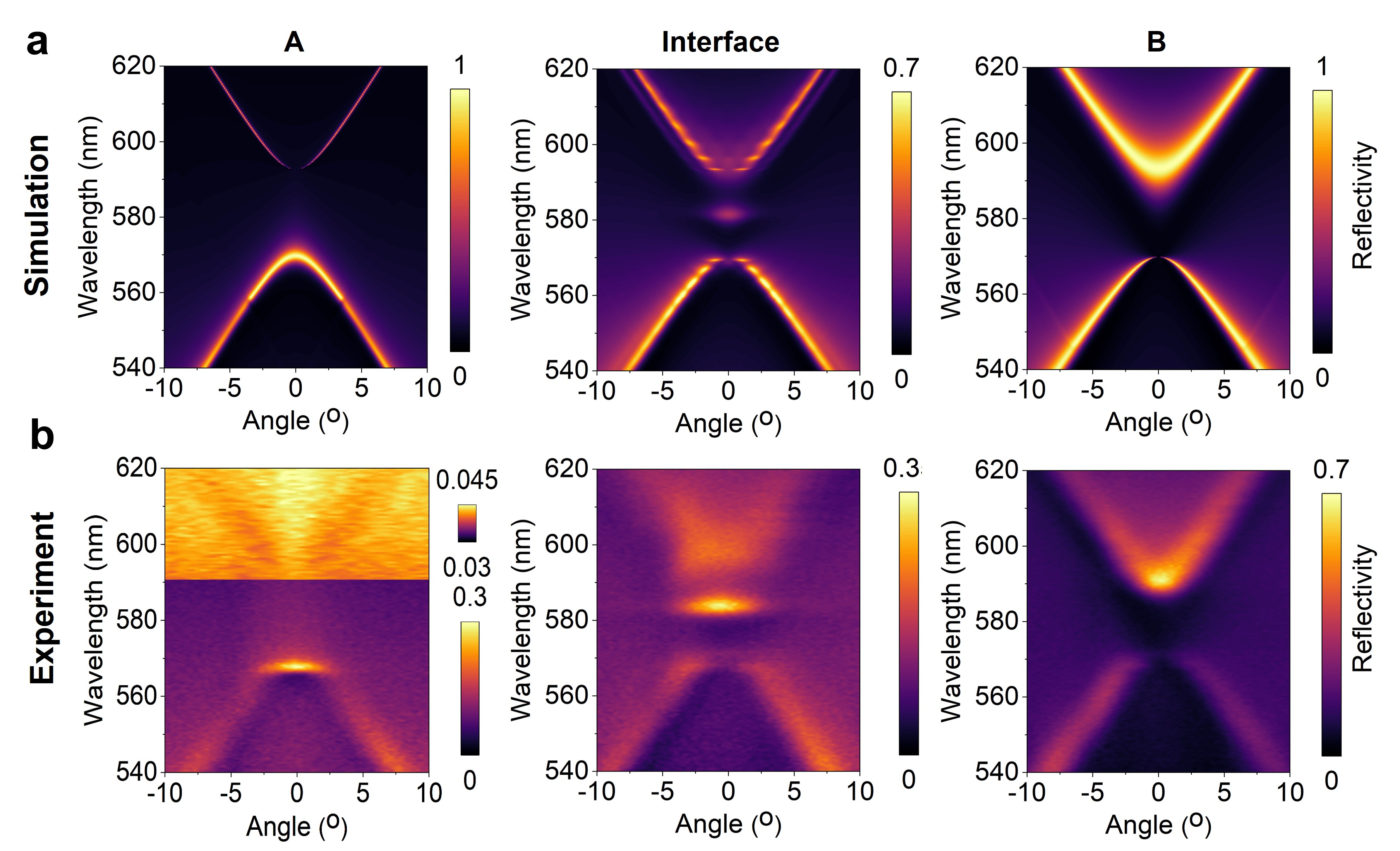}
    \caption{Reflection for single JR. (a) Numerical simulation for grating A, interface and grating B. (b) Experimental reflection for grating A, interface and grating B.}
    \label{s3}
\end{figure*}

\begin{figure*}[htb!]
    \centering
    \includegraphics[width=0.5\linewidth]{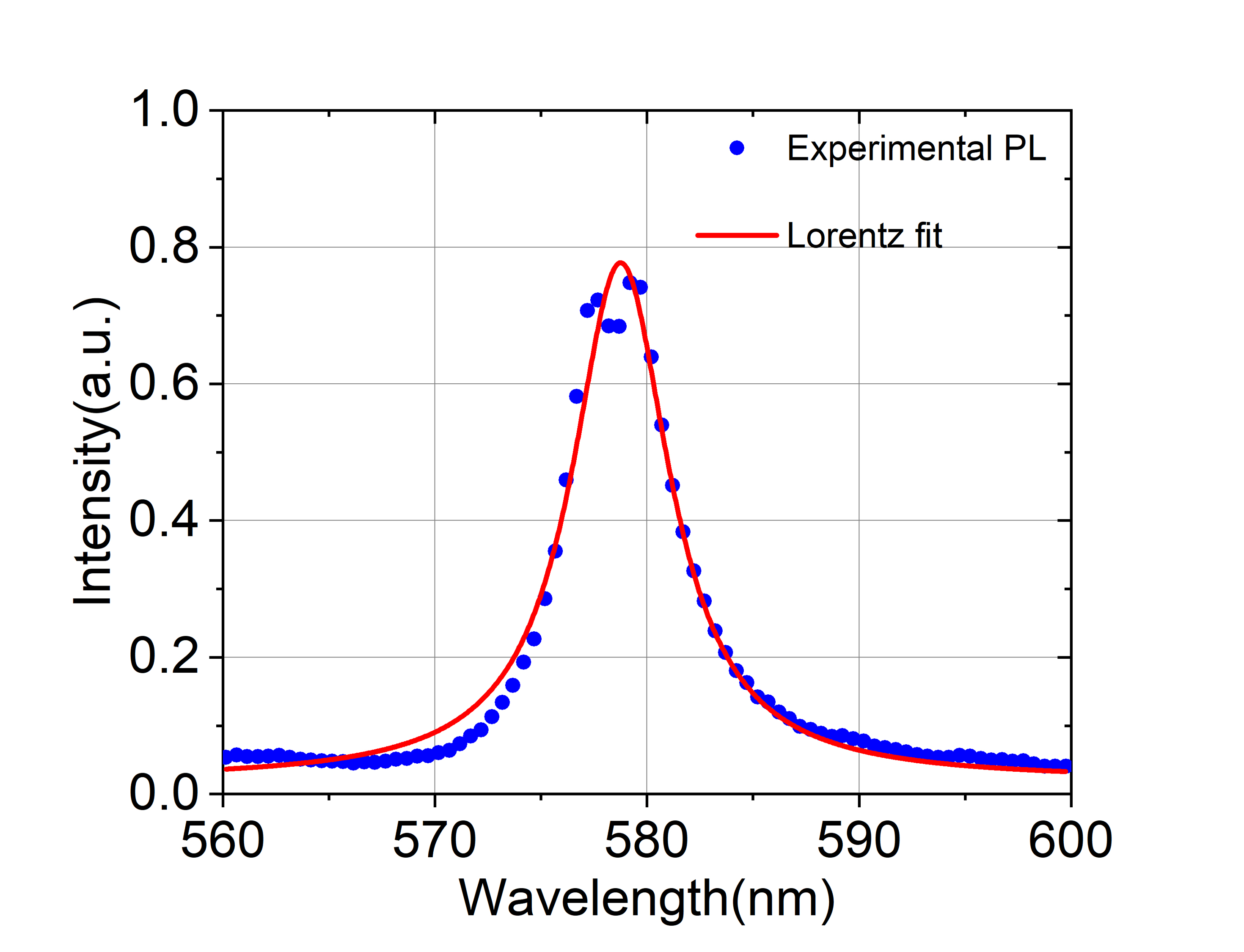}
    \caption{PL spectra for Nile red mixed in PMMA on the SiO\(_2\) substrate.}
    \label{s4}
\end{figure*}

\begin{figure*}[htb!]
    \centering
    \includegraphics[width=0.5\linewidth]{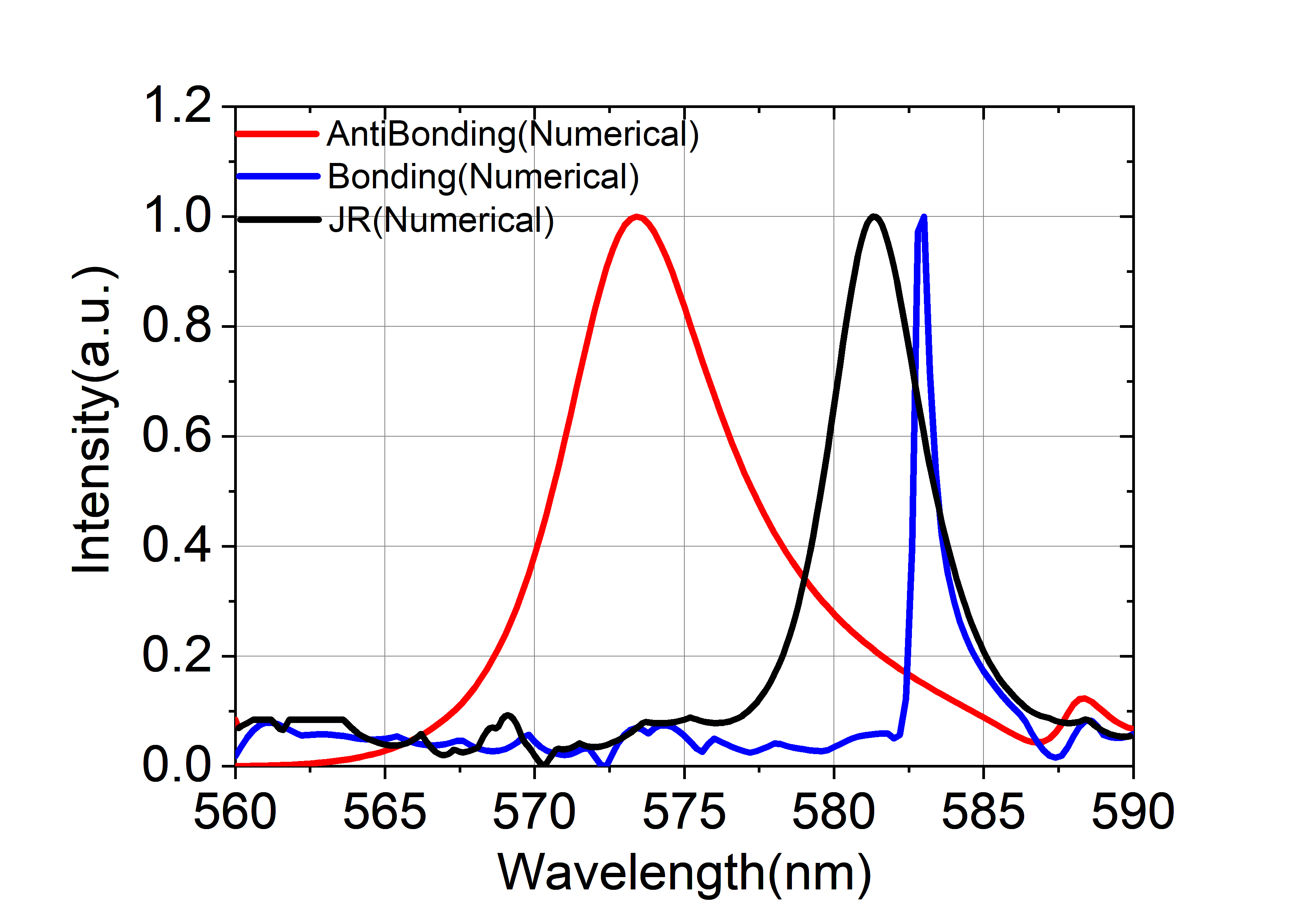}
    \caption{Numerical simulation of PL for JR, bonding and anti-bonding modes.}
    \label{s5}
\end{figure*}

\begin{figure*}[htb!]
    \centering
    \includegraphics[width=1\linewidth]{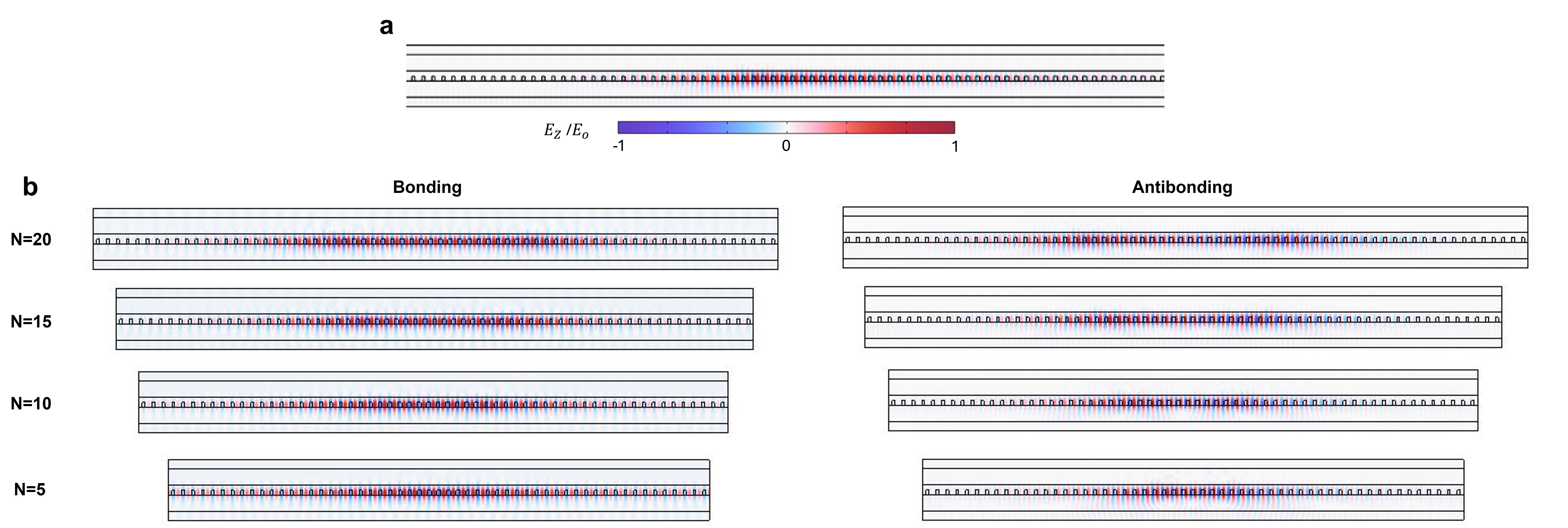}
    \caption{Evolution of electric field distribution respecting to the chnage of separation distance (grating B cells number N). (a) JR mode distribution. (b) mode distribution of bonding and anti-bonding states with different separation distance (N=20,15,10,5)}
    \label{s6}
\end{figure*}

\begin{figure*}[htb!]
    \centering
    \includegraphics[width=0.8\linewidth]{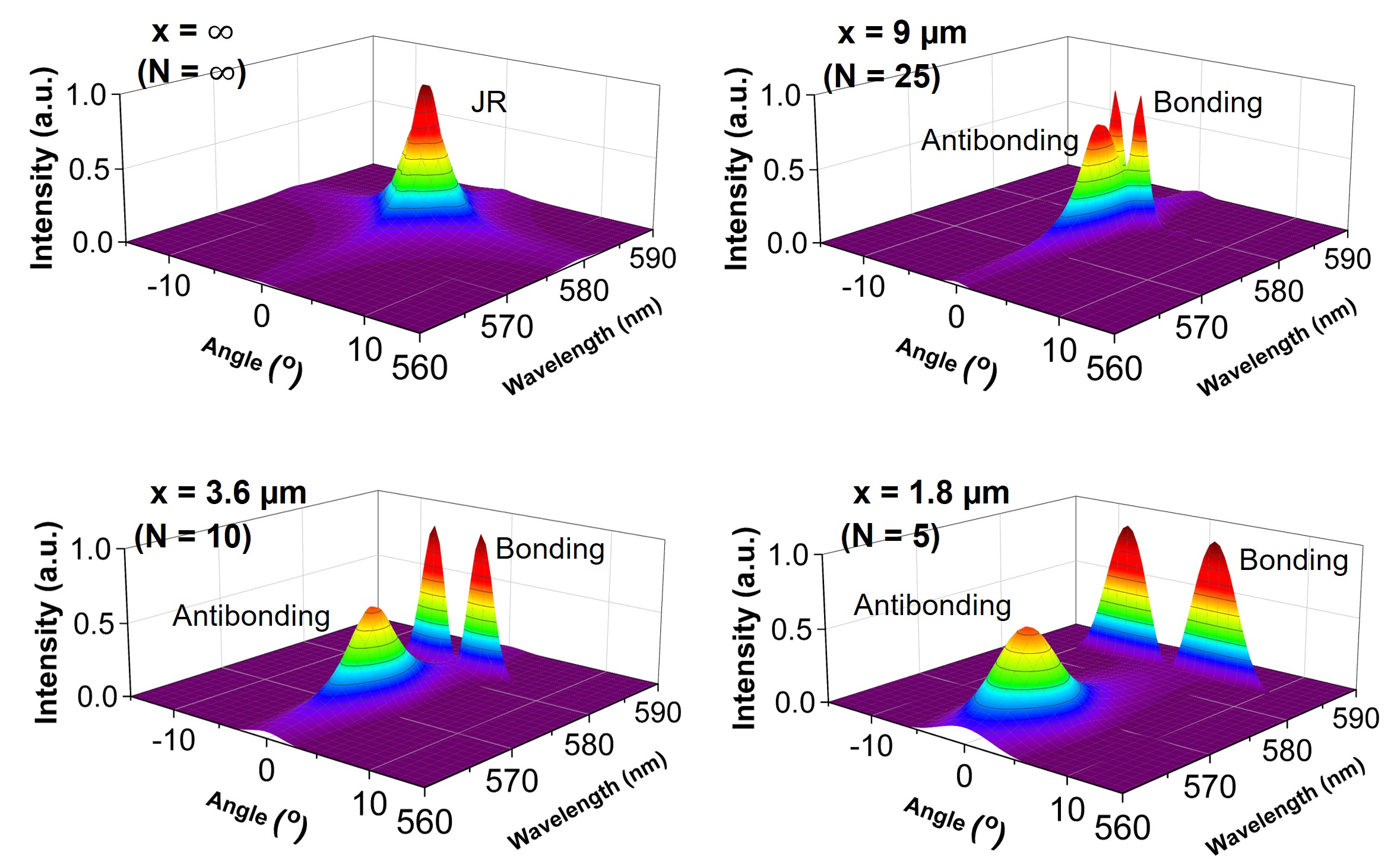}
    \caption{Emission spectra of the hybrid JR system, calculated by the analytical model for different sizes of grating B.}
    \label{s7}
\end{figure*}

\begin{figure*}[htb!]
    \centering
    \includegraphics[width=0.4\linewidth]{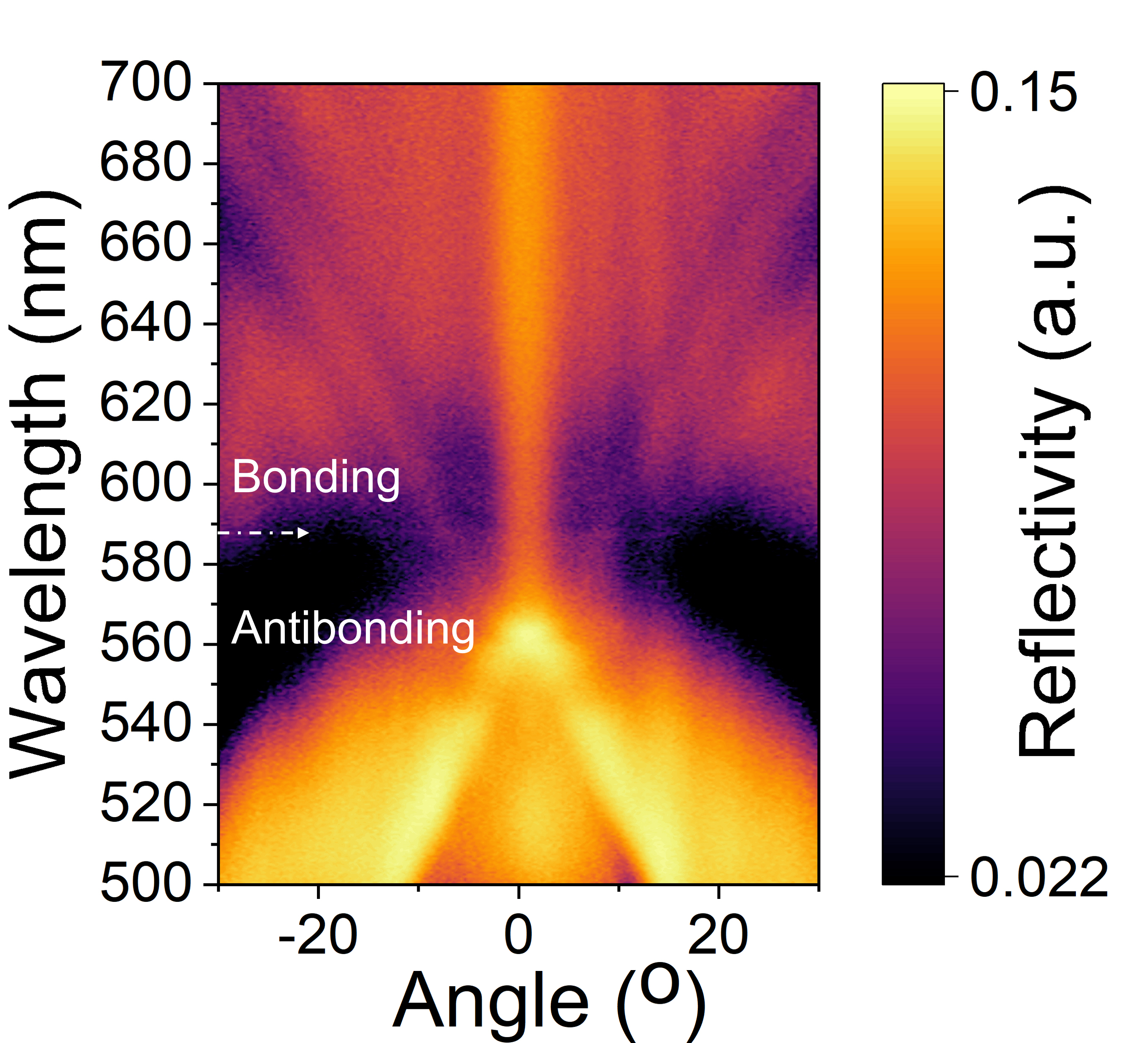}
    \caption{Reflection for double-junction when N=5. The Bonding mode is located at 587nm while the anti-bonding mode is at 563nm.}
    \label{s8}
\end{figure*}

\end{document}